\documentclass[conference]{IEEEtran}
\IEEEoverridecommandlockouts

\usepackage{fancyhdr}
\fancypagestyle{cfooter}{ %
\fancyhf{} 
\cfoot{\scriptsize{This work has been submitted to the IEEE for possible publication. Copyright may be transferred without notice, after which this version may no longer be accessible.}
}

}

\usepackage{cite}
\usepackage{amsmath,amssymb,amsfonts}
\usepackage{algorithmic}
\usepackage{graphicx}
\usepackage{textcomp}
\usepackage{xcolor}
\usepackage{url}

\usepackage{cite}
\usepackage{amsmath,amssymb,amsthm,fixmath}
\usepackage{mathtools}
\usepackage{xcolor}
\usepackage{bm}
\usepackage{makecell}
\usepackage{pgfplots}
\usepackage{fancyhdr}
\usepackage{hyperref}
\usepackage{caption}
\usepackage{xcolor}
\usepackage{bm}
\usepackage{makecell}
\usepackage{pgfplots}
\usepackage{tikz}
\usepackage{graphicx}
\usepackage{booktabs} 
\usetikzlibrary{intersections,pgfplots.fillbetween,patterns,spy,shapes.geometric,calc,positioning}
\usetikzlibrary{arrows,decorations.pathmorphing,backgrounds,fit,matrix,fillbetween}
\usetikzlibrary{shapes.geometric, arrows.meta}
\usetikzlibrary{decorations.pathreplacing,decorations.markings}
\usepgfplotslibrary{groupplots}
\usepackage{soul}
\usepackage{url}
\usepackage{float}
\usepackage{adjustbox}
\usetikzlibrary {arrows.meta}
\usepackage{acro}
\usepackage{balance}

\DeclareMathOperator{\E}{\mathbb{E}}

\DeclareAcronym{FDD}{short = FDD ,long = frequency-division duplexing}
\DeclareAcronym{CS}{short = CS ,long = compressive sensing}
\DeclareAcronym{AE}{short = AE ,long = autoencoder}
\DeclareAcronym{PCA}{short = PCA ,long = principal component analysis}
\DeclareAcronym{DM}{short = DM ,long = diffusion model}
\DeclareAcronym{CIR}{short = CIR ,long = channel impulse response}
\DeclareAcronym{MMSE}{short = MMSE,long =  minimum mean squared error}
\DeclareAcronym{MSE}{short = MSE, long = mean squared error}
\DeclareAcronym{NMSE}{short = NMSE, long = normalized MSE}
\DeclareAcronym{LMMSE}{short = LMMSE, long = linear minimum mean squared error}
\DeclareAcronym{WSS}{short = WSS, long = wide sense stationary}
\DeclareAcronym{DFT}{short = DFT, long = discrete Fourier transform}
\DeclareAcronym{FT}{short = FT, long = Fourier transform}
\DeclareAcronym{EM}{short = EM, long = expectation-maximization}
\DeclareAcronym{BIC}{short = BIC, long = Bayesian information criterion}
\DeclareAcronym{AIC}{short = AIC, long = Akaike information criterion}
\DeclareAcronym{i.i.d.}{short = i.i.d., long = independent and identically distributed}
\DeclareAcronym{FLOP}{short = FLOP, long = floating point operation}
\DeclareAcronym{AWGN}{short = AWGN, long = additive white Gaussian noise}
\DeclareAcronym{LS}{short = LS, long = least squares}
\DeclareAcronym{ARMA}{short = ARMA, long = autoregressive moving-average}
\DeclareAcronym{FBM}{short = FBM, long = fractional Brownian motion}
\DeclareAcronym{FFT}{short = FFT, long = fast Fourier transform}
\DeclareAcronym{GMM}{short = GMM, long = Gaussian mixture model}
\DeclareAcronym{VAE}{short = VAE, long = variational autoencoder}
\DeclareAcronym{NN}{short = NN, long = neural network}
\DeclareAcronym{PD}{short = PD, long = positive definite}
\DeclareAcronym{OP}{short = OP, long = optimization problem}
\DeclareAcronym{GS}{short = GS, long = Gohberg-Semencul}
\DeclareAcronym{CSI}{short = CSI, long = channel state information}
\DeclareAcronym{SAVG}{short = SAVG, long = SCM averaged along its diagonals}
\DeclareAcronym{UE}{short = UE, long = user equipment}
\DeclareAcronym{DoA}{short = DoA, long = direction of arrival}
\DeclareAcronym{DoAs}{short = DoAs, long = directions of arrival}
\DeclareAcronym{DoDs}{short = DoDs, long = directions of depature}
\DeclareAcronym{ML}{short = ML, long = machine learning}
\DeclareAcronym{BS}{short = BS, long = base station}
\DeclareAcronym{WSSUS}{short = WSSUS, long = wide-sense-stationary-uncorrelated-scattering}
\DeclareAcronym{ULA}{short = ULA, long = uniform linear array}
\DeclareAcronym{URA}{short = URA, long = uniform rectangular array}
\DeclareAcronym{MIMO}{short = MIMO, long = multiple-input-multiple-output}
\DeclareAcronym{SIMO}{short = SIMO, long = single-input-multiple-output}
\DeclareAcronym{OFDM}{short = OFDM, long = orthogonal-frequency-division-multiplexing}
\DeclareAcronym{mmWave}{short = mmWave,long =  millimeter wave}
\DeclareAcronym{HST}{short = HST, long = high-speed train}
\DeclareAcronym{UAV}{short = UAV, long = unmanned aerial vehicles}
\DeclareAcronym{IoT}{short = IoT, long = Internet of things}
\DeclareAcronym{Eig}{short = Eig, long = eigenvalue}
\DeclareAcronym{Frob}{short = Frob, long = Frobenius}
\DeclareAcronym{PGD}{short = PGD, long = projected gradient descent}
\DeclareAcronym{PLS}{short = PLS, long = projected LS}
\DeclareAcronym{LOS}{short = LOS,  long =line-of-sight}
\DeclareAcronym{GAN}{short = GAN, long = generative adversarial network}
\DeclareAcronym{NLOS}{short = NLOS, long = non-line-of-sight}
\DeclareAcronym{SISO}{short = SISO, long = single-input-single-output}
\DeclareAcronym{SNR}{short = SNR, long = signal-to-noise ratio}
\DeclareAcronym{BN}{short = BN, long = Bayesian network}
\DeclareAcronym{KL}{short = KL, long = Kullback-Leibler}
\DeclareAcronym{ELBO}{short = ELBO, long = evidence lower bound}
\DeclareAcronym{CGLM}{short = CGLM, long = conditionally Gaussian latent model}
\DeclareAcronym{PGMM}{short = PGMM, long = parameter GMM}
\DeclareAcronym{PVAE}{short = PVAE, long = parameter VAE}
\DeclareAcronym{AP}{short = AP, long = access point}
\DeclareAcronym{PMF}{short = PMF, long = probability mass function}

\DeclareAcronym{CDL}{short = CDL,long = cluster delay line}
\DeclareAcronym{TDL}{short = TDL,long = tap delay line}
\DeclareAcronym{EPA}{short = EPA,long = extended pedestrian A}
\DeclareAcronym{GSCM}{short = GSCM,long = geometry-based stochastic channel model}
\DeclareAcronym{CME}{short = CME,long = conditional mean estimator}

\DeclareAcronym{CSGMM}{short = CSGMM,long = compressive sensing \ac{GMM}}
\DeclareAcronym{CSVAE}{short = CSVAE,long = compressive sensing \ac{VAE}}
\DeclareAcronym{CP-GMM}{short = CP-GMM,long = channel parameter-\ac{GMM}}
\DeclareAcronym{CP-VAE}{short = CP-VAE,long = channel parameter-\ac{VAE}}
\DeclareAcronym{SBL}{short = SBL,long = sparse Bayesian learning}
\DeclareAcronym{SBGM}{short = SBGM, long  = sparse Bayesian generative modeling}
\DeclareAcronym{PG}{short = PG, long  = path gain}
\DeclareAcronym{GM}{short = GM, long  = Generative model}
\DeclareAcronym{DL}{short = DL, long  = deep learning}
\DeclareAcronym{LLM}{short = LLM, long  = large language model}
\DeclareAcronym{JCAS}{short = JCAS, long  = joint communication and sensing}
\DeclareAcronym{RMSE}{short = RMSE, long  = root mean squared error}
\DeclareAcronym{CDF}{short = CDF, long  = cumulative density function}

\definecolor{color1}{rgb}{0 0.4470 0.7410}
\definecolor{color2}{rgb}{0.3010 0.7450 0.9330}
\definecolor{color3}{rgb}{0.4940 0.1840 0.5560}
\definecolor{color4}{rgb}{0.6350 0.0780 0.1840}
\definecolor{color5}{rgb}{0.8500 0.3250 0.0980}
\definecolor{color6}{rgb}{0.9290 0.6940 0.1250}
\definecolor{color7}{rgb}{0 0.50 0}
\definecolor{color8}{rgb}{0.4660 0.6740 0.1880}
	
\definecolor{Yellow}{rgb}{1.00, 0.71, 0.00}
\definecolor{Orange}{rgb}{1.00, 0.50, 0.00}
\definecolor{Red}{rgb}{0.90, 0.20, 0.09}
\definecolor{DarkRed}{rgb}{0.9, 0.3, 0.3}
\definecolor{Blue}{rgb}{0.00, 0.60, 1.00}
\definecolor{LightBlue}{rgb}{0.25, 0.75, 1.00}
\definecolor{Green}{rgb}{0.57, 0.67, 0.42}
\definecolor{LightGreen}{rgb}{0.71, 0.89, 0.58}
\definecolor{Black}{rgb}{0,0,0}
\definecolor{Gray}{rgb}{0.7,0.7,0.6}
\definecolor{DarkGray}{rgb}{0.4,0.4,0.3}
\definecolor{color1bg}{HTML}{ECD9ED}
	
\definecolor{MidnightBlue}{HTML}{006795}
\definecolor{SpringGreen}{HTML}{C6DC67}
\definecolor{PineGreen}{HTML}{008B72}
\definecolor{Maroon}{HTML}{AF3235}
\definecolor{RedOrange}{HTML}{F26035}
\definecolor{SkyBlue}{HTML}{46C5DD}
\definecolor{Dandelion}{HTML}{F7921D}
\definecolor{Periwinkle}{HTML}{7977B8}
\definecolor{RedViolet}{HTML}{3C8031}

\definecolor{asparagus}{rgb}{0.53, 0.66, 0.42}
\definecolor{burntsienna}{rgb}{0.91, 0.45, 0.32}
\definecolor{cadetblue}{rgb}{0.37, 0.62, 0.63}
\definecolor{carminepink}{rgb}{0.92, 0.3, 0.26}
\usetikzlibrary{external}
\def\BibTeX{{\rm B\kern-.05em{\sc i\kern-.025em b}\kern-.08em
    T\kern-.1667em\lower.7ex\hbox{E}\kern-.125emX}}
\begin{document}
\title{Wireless Channel Modeling for Machine Learning ‑ A Critical View on Standardized Channel Models \\
}

\author{\IEEEauthorblockN{Benedikt Böck, Amar Kasibovic, and Wolfgang Utschick}
\IEEEauthorblockA{\textit{TUM School of Computation, Information and Technology, Technical University of Munich, Germany} \\
\{benedikt.boeck, amar.kasibovic, utschick\}@tum.de}
}

\maketitle

\thispagestyle{cfooter}

\begin{abstract}
Standardized (link-level) channel models such as the 3GPP TDL and CDL models are frequently used to evaluate \ac{ML}-based physical-layer methods. However, in this work, we argue that a link-level perspective incorporates limiting assumptions, causing unwanted distributional shifts or necessitating impractical online training. 
An additional drawback is that this perspective leads to (near-)Gaussian channel characteristics. Thus, \ac{ML}-based models, trained on link-level channel data, do not outperform classical approaches for a variety of physical-layer applications. Particularly, we demonstrate the optimality of simple linear methods for channel compression, estimation, and modeling, revealing the unsuitability of link-level channel models for evaluating ML models. On the upside, adopting a scenario-level perspective offers a solution to this problem and unlocks the relative gains enabled by \ac{ML}.


\end{abstract}


\section{Introduction}

The application of \ac{ML} requires access to datasets from which parameterized models can learn. While most research in \ac{ML} focuses on advancing the state-of-the-art techniques for extracting the desired information encoded in the dataset, the careful design of the dataset itself is foundational for the model to work properly. Arguably, the introduction of benchmark datasets such as MNIST and ImageNet has led to key breakthroughs in \ac{ML} for imaging  \cite{paullada2021}. 
Compared to imaging, where ground-truth data is rather easily accessible, other domains, such as wireless communications, exhibit additional challenges for a proper design of datasets. 
A general desire for datasets in wireless communications is to be site-specific, i.e., all data originates from one particular propagation environment \cite{Alkhateeb2019}. This is due to the dependence of the wireless channel on specific environmental factors such as the scatterers' positions and surface material. Consequently, site-specific channel data can be used to train \ac{ML} models that exhibit superior performance in that same environment compared to generically trained models. However, acquiring site-specific high-quality measurement datasets for every environment of interest is prohibitively costly and time-consuming. One promising alternative is to rely on channel models capable of generating synthetic yet realistic site-specific channel data.

Channel modeling has been investigated well before \ac{ML} has emerged for wireless communications \cite{bello1963}. Early channel models, such as the one-ring model, assume simplistic scatterer positions and an infinite number of paths to determine the small-scale fading characteristics at a single position \cite{Clarke1968}. In the past years, advanced geometry-based stochastic channel models such as QuaDRiGa \cite{Jaeckel2014} or the COST \cite{Cost2100} channel models have been introduced and simulate channel realizations based on a random placement of scatterers. Other channel models are based on ray tracing techniques, which determine channel realizations in a fully deterministic manner using a 3D scene and the material properties of scatterers. \cite{orekondy2023winert}. Another branch of modern channel models is based on generative modeling. Early work utilizes \acp{GAN} to learn the channel distribution \cite{Yang2019}. Recent work \cite{bock2025physicsinformed} addresses the drawbacks of \acp{GAN} by combining \ac{SBGM} \cite{boeck2024nips} with model-based insights of conditional channel statistical moments \cite{boeck2024wcl}. This decades-long research has led to a landscape of different channel models from which many have never been intended to provide \ac{ML} training data, raising the question, which model should be used and avoided in the context of \ac{ML}. 

In this work, we critically review standardized link-level channel models comprising the 3GPP TDL A-E and CDL A-E models \cite{3gpptdlcdl}, as well as the (extended) pedestrian and vehicular A and B models \cite{3gppEPA}. These models are frequently used for evaluating \ac{ML}-based methods, mainly due to their easy-to-use implementation in the \textit{5G Toolbox} and \textit{LTE Toolbox} of MATLAB. 
However, since these models have not been intended for \ac{ML}, they incorporate very restrictive assumptions that, to the best of our knowledge, have not been rigorously discussed in the literature. 
In our opinion, these assumptions have serious implications for the meaningfulness of experimental results based on these models. 
Our main contributions are as follows:
\begin{itemize}
    \item By adopting a link- vs. scenario-level perspective, we argue that training \ac{ML} models based on link-level models' channel data generally incurs unwanted distributional shifts or necessitates impractical online training.
    \item We show that link-level models' channel realizations are (near-)Gaussian, making linear methods provably superior for many physical-layer applications, with \ac{ML} models largely reduced to approximating this linear mapping.
    \item We illustrate this property with three examples. The \ac{LMMSE} estimator outperforms \ac{ML} models for channel estimation, the \ac{PCA} outperforms \acp{AE} for \ac{CSI} compression, and Gaussian sampling via the sample covariance outperforms generative models for channel generation. 
\end{itemize}
Thus, simulation results obtained from link-level channel models offer limited to no evidence regarding the true performance of an \ac{ML}-based method in practical wireless settings. Interestingly, adapting a scenario-level perspective and using advanced channel models such as QuaDRiGa overcomes all the mentioned limitations. Thus, we strongly advocate this perspective, as it is practically reasonable and simultaneously enables \ac{ML}-based methods to surpass the performance of classical signal processing schemes. 

\section{Link-Level vs. Scenario-Level Perspective}

\begin{figure}[t]
    \centering
\includegraphics{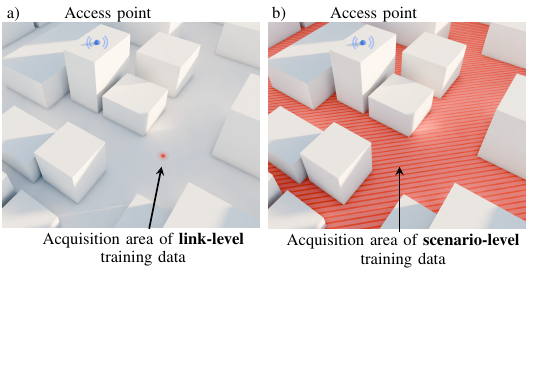}
   \vspace{-2.5cm}
\caption{Link-level vs scenario-level perspective.}
\label{fig:linklevelvescenariolevel}
	\vspace{-0.7cm}
\end{figure}

Standardized link-level channel models such as the 3GPP TDL and CDL models assume path powers, path delays, and path angles to be fixed \cite{3gpptdlcdl}. The only variation between realizations drawn from these channel models lies in the fading coefficients of individual paths. 
As an example, the TDL-A model assumes the static \ac{CIR} as
\begin{equation}
    \label{eq:tdl_a_cir}
    h(\tau) = \sum_{l=1}^L \sqrt{p_l}e_l\delta(\tau - \tau_l)
\end{equation}
with fixed constant path powers $p_l$, path delays $\tau_l$ and varying fading coefficients $e_l \sim \mathcal{N}_{\mathbb{C}}(0,1)$. These fading coefficients model the small-scale characteristics of the channel that occur when a user moves over a distance of a few wavelengths. However, by moving further, $\tau_l$ and $p_l$ are changing variables, which is not reflected by the link-level model. In consequence, generating channel realizations from \eqref{eq:tdl_a_cir} is equivalent to acquiring a training dataset from a single fixed position and its small-scale fading area spanning a few wavelengths in a real wireless scenario. This setup is illustrated in Fig. \ref{fig:linklevelvescenariolevel} a). 

From an \ac{ML} perspective, this implies that when a user moves outside this specific area, the link-level trained model encounters a distributional shift between training and test data, and its performance depends on its robustness against such shifts. One way to address this issue is by incorporating online training. This approach, however, requires frequent acquisition of training data as well as repeated model training, which is questionable considering the strict latency- and compute-restrictions in wireless communications. When sampling the \ac{CIR} in \eqref{eq:tdl_a_cir} and stacking the resulting entries in a vector $\bm{h}$, we can interpret link-level channel models as conditional channel models that are conditioned on the path powers, delays, and angles summarized in a vector $\bm{\delta}$. Thus, the training dataset is given by $\{\bm{h}_i | \bm{h}_i \sim p(\bm{h}|\bm{\delta})\}_{i=1}^N$ for one fixed $\bm{\delta}$.

The alternative scenario-level perspective assumes training channel data with varying path powers, path angles, and path delays. By adapting the conditional channel modeling perspective of link-level models, we can interpret the generation of channel realizations $\bm{h}_i$ in scenario-level channel models as 
\begin{equation}
    \bm{h}_i \sim p(\bm{h}) = \int p(\bm{h}|\bm{\delta})p(\bm{\delta})\text{d}\bm{\delta}.
\end{equation}
Thus, every scenario-level channel model contains a link-level channel model together with a parameter prior $p(\bm{\delta})$. Since $\bm{\delta} $ is allowed to change when sampling, this perspective is equivalent to acquiring a training dataset from the whole environment that is served by an \ac{AP} and is illustrated in Fig. \ref{fig:linklevelvescenariolevel} b). The important benefit of this perspective is that as long as a user does not move outside the area from which training data has been acquired, there is no distributional shift between test and training data and no need for online training. 

\section{Gaussianity of Link-Level Standardized Channel Models}

One might intuitively argue that the link-level perspective remains reasonable, since an \ac{ML}-based method should perform better when training and testing are restricted to a subset of the data, potentially making the extra effort of online training or robustification worthwhile. For example, an MNIST digit classifier achieves higher accuracy when distinguishing only digits one to three than when covering zero to nine. Perhaps surprisingly, for many physical-layer applications, the opposite holds true when considering the relative performance gains over classical methods. Only with a scenario-level perspective can one significantly benefit from \ac{ML}-based methods. This is due to the (near-)Gaussianity of standardized link-level channel models, for which simple linear schemes, easily derived from the given training data, are optimal for many physical-layer applications. In the following, we rigorously validate the (near-)Gaussianity of the mostly used link-level channel models, when using them for generating \ac{OFDM} or \ac{MIMO} channel data.

\subsection{Tapped Delay Line Models}
\label{sec:tdl}
As described in \eqref{eq:tdl_a_cir}, the TDL models assume fixed path powers and delays. In general, they also allow for time-varying fading coefficients, which are modeled using the Jakes spectrum. The TDL-D and TDL-E models also incorporate a \ac{LOS} path with slightly unique characteristics, which is why we first restrict ourselves to the TDL A-C models. There, the time varying \ac{CIR} is given by \cite{3gpptdlcdl}
\begin{equation}
    \label{eq:tdl_a_cir_tv1}
    h(\tau; t) = \sum_{l=1}^L \sqrt{p_l}e_l(t)\delta(\tau - \tau_l)
\end{equation}
with $e_l(t)$ being \ac{i.i.d.} Gaussian processes with mean $\E[e_l(t)] = 0$ and autocovariance $\E[e_l(t)\overline{e_l(t-\eta)}] = \operatorname{J}_0(2\pi f_D \eta)$. Here, $\operatorname{J}_0(\cdot)$ denotes the zeroth-order Bessel function and $f_D$ is the maximum Doppler shift. To end up with \ac{OFDM} channel representations, we transform \eqref{eq:tdl_a_cir_tv1} with respect to $\tau$ to the frequency domain via the Fourier transform and sample the resulting expression equidistantly with subcarrier spacing $\Delta f$ and symbol duration $\Delta T$. This results in the \ac{OFDM} channel matrix
\begin{equation}
    \label{eq:ofdm_channel_matrix}
    \bm{H} = \sum_{l=1}^L \sqrt{p_l} \bm{a}_t^{(l)}\bm{a}_f^{(l)\operatorname{T}}
\end{equation}
with  $\bm{a}_f^{(l)} = [1,\operatorname{e}^{-\operatorname{j}2\pi\Delta f \tau_l},\ldots,\operatorname{e}^{-\operatorname{j}2\pi\Delta f (N_f-1) \tau_l}]^{\operatorname{T}} \in \mathbb{C}^{N_f}$ and 
$\bm{a}_t^{(l)} \sim \mathcal{N}_{\mathbb{C}}(\bm{0},\bm{C}_{\text{Jakes}})$ with $\bm{C}_{\text{Jakes}}\large|_{ij} = \operatorname{J}_0(2\pi f_D \Delta T |i-j|)$ for all $l$. Vectorizing \eqref{eq:ofdm_channel_matrix} leads to 
\begin{equation}
    \label{eq:ofdm_channel_matrix}
    \bm{h} = \text{vec}(\bm{H}) = \sum_{l=1}^L \sqrt{p_l} \left(\bm{a}_t^{(l)} \otimes \bm{a}_f^{(l)}\right).
\end{equation}
It can be directly inferred that due to $\tau_l$ and $p_l$ being constants and not random variables, $\bm{h}$ follows a Gaussian distribution as it is a linear combination of Gaussians $\bm{a}_t^{(l)}$. The channel $\bm{h}$ has a zero mean due to $\E[\bm{a}_t^{(l)}] = \bm{0}$ for all $l$. Moreover, the channel's covariance $\bm{C}_{\text{TDL}}$ is given by
\begin{equation}
    \bm{C}_{\text{TDL}} = \E[\bm{h}\bm{h}^{\operatorname{H}}] = \sum_l p_l \left( \bm{C}_{\text{Jakes}} \otimes \bm{a}_f^{(l)}\bm{a}_f^{(l)\operatorname{H}} \right)
\end{equation}
where we utilized that $\E[\bm{a}_t^{(l)}\bm{a}_t^{(m)\operatorname{H}}] = \delta_{m-l}\bm{C}_{\text{Jakes}}$ and the commutation of the Kronecker and matrix multiplication.

In addition to the description in \eqref{eq:tdl_a_cir_tv1}, the TDL-D and TDL-E models contain a further \ac{LOS} path whose fading coefficient has a uniform phase, but a fixed absolute value and an additional peak in its spectrum. Consequently, neither of the latter two models is perfectly Gaussian but rather near-Gaussian.

\subsection{Clustered Delay Line Models} 
\label{sec:cdl}
Compared to 3GPP TDL models, CDL models incorporate the additional distinction between clusters (i.e., main paths) and sub-paths. Specifically, when a ray is reflected by an obstacle, diffuse reflection may occur, where the roughness of the reflecting surface can produce multiple rays with similar characteristics. Consequently, a distinction must be made between the cluster (or main path) properties and the properties of the individual rays (or sub-paths). While the CDL models incorporate delays, we analyze their assumptions in the spatial domain, as the derivation for the delay is equivalent to the one for TDL models (cf. Section \ref{sec:tdl}). We first restrict ourselves to the CDL-A to C models where the \ac{LOS} path is absent.

In general, the CDL models allow for customizing the radiation pattern as well as the antenna polarization.
The generic CDL \ac{MIMO} channel matrix equals \cite{3gpptdlcdl}
\begin{equation}
\label{eq:cdl_mimo}
    \bm{H}\hspace{-0.05cm} = \hspace{-0.05cm} \sum_{l=1}^{L} \frac{1}{\sqrt{M}} \sum_{m=1}^M \rho_{lm}\bm{a}_{\text{rx}}(\phi_{\text{rx}}^{(lm)}, \theta_{\text{rx}}^{(lm)})\bm{a}_{\text{tx}}(\phi_{\text{tx}}^{(lm)}, \theta_{\text{tx}}^{(lm)})^{\operatorname{T}}
\end{equation}
with azimuth and elevation angles $\phi_{(\cdot)}^{(lm)}$, $\theta_{(\cdot)}^{(lm)}$ of the $m$th sub-path of cluster $l$. The vectors $\bm{a}_{\text{rx}}(\cdot, \cdot)$ and $\bm{a}_{\text{tx}}(\cdot, \cdot)$ are the so-called steering vectors and encode the antennas' spatial positions $\bm{d}^{(i)}_{\text{rx}}$ and $\bm{d}^{(j)}_{\text{tx}}$ in local coordinate systems, i.e., 
\begin{equation}
    \bm{a}_{(\cdot)}(\phi_{(\cdot)}^{(lm)}, \theta_{(\cdot)}^{(lm)})\big|_i = \operatorname{e}^{\operatorname{j}\frac{2\pi}{\lambda}\bm{e}(\phi_{(\cdot)}^{(lm)}, \theta_{(\cdot)}^{(lm)})^{\operatorname{T}}\bm{d}^{(i)}_{(\cdot)}}
\end{equation} 
with wavelength $\lambda$ and spherical unit vector $\bm{e}(\cdot, \cdot)$. For the considerations in this work, we can summarize the effects of the real-valued path loss, the cross polarization, random path phase shifts, as well as the polarization and radiation pattern of receiving and transmitting antenna arrays in one single variable $\frac{1}{\sqrt{M}}\rho_{l,m}$, usually referred to as complex path loss \cite{bock2025physicsinformed}. One important feature for what follows is that the phases $\beta_{lm}$ of $\rho_{lm} = |\rho_{lm}|\operatorname{e}^{\operatorname{j}\beta_{lm}}$ are generally \ac{i.i.d.} with $\beta_{lm} \sim \mathcal{U}(-\pi,\pi)$. CDL models assume $\phi_{(\cdot)}^{(lm)}$, $\theta_{(\cdot)}^{(lm)}$ and most effects contained in $\rho_{lm}$ to be fixed constants. The only variation between realizations drawn from these channel models lies in the random phases $\beta_{lm}$. 

While $\phi_{(\cdot)}^{(lm)}$, $\theta_{(\cdot)}^{(lm)}$ are constants, they are chosen in a specific manner, such that they statistically resemble \ac{i.i.d.} samples drawn from Laplacian distributions parameterized by fixed cluster angles, i.e., 
\begin{equation}
\label{eq:iid_angles}
    \begin{aligned}
        \phi_{(\cdot)}^{(lm)} \sim p(\phi ; \bar{\phi}_{(\cdot)}^{(l)}) = \mathrm{Laplacian}(\bar{\phi}_{(\cdot)}^{(l)}, c^{(\phi)}_{(\cdot)}) \\
         \theta_{(\cdot)}^{(lm)} \sim p(\theta ; \bar{\theta}_{(\cdot)}^{(l)}) = \mathrm{Laplacian}(\bar{\theta}_{(\cdot)}^{(l)}, c^{(\theta)}_{(\cdot)}) \\
    \end{aligned}
\end{equation}
with cluster angles $\bar{\phi}_{(\cdot)}^{(l)}$ and $\bar{\theta}_{(\cdot)}^{(l)}$ and $\mathrm{Laplacian}(\mu, b)$ being the Laplacian with mean $\mu$ and spread $b$. The number of sub-paths $M$ in \eqref{eq:cdl_mimo} is generally chosen to be $20$. These considerations allow us to apply the central limit theorem to the inner summation in \eqref{eq:cdl_mimo}. After vectorizing \eqref{eq:cdl_mimo}, we define
\begin{equation}
    \bm{h}_{lm} = \rho_{lm} \left( \bm{a}_{\text{rx}}^{(lm)} \otimes \bm{a}_{\text{tx}}^{(lm)} \right).
\end{equation}
with \ac{i.i.d.} random vectors $\bm{h}_{lm}$.\footnote{For simplicity, we abbreviated the notation for the steering vectors.} Thus, the central limit theorem allows us to approximate 
\begin{equation}
    \label{eq:central_limit}
    \frac{1}{\sqrt{M}} \sum_{m=1}^M \bm{h}_{lm} \sim \mathcal{N}_{\mathbb{C}}(\bm{0}, \bm{C}_l).
\end{equation}
The zero mean stems from $\beta_{lm} \sim \mathcal{U}(-\pi,\pi)$. Moreover, 
\begin{equation}
    \bm{C}_l = \E\left[ |\rho_{lm}|^2  \left(\bm{a}_{\text{rx}}^{(lm)}\bm{a}_{\text{rx}}^{(lm)\operatorname{H}} \right) \otimes \left(\bm{a}_{\text{tx}}^{(lm)}\bm{a}_{\text{tx}}^{(lm)\operatorname{H}}\right) \right].
\end{equation}
where we used that $\E[\rho_{lm}\rho_{k,f} \cdot C] = \E[|\rho_{lm}|^2  \cdot C]\delta_{l-k}\delta_{m-f}$ for any random variable $C$ and the commutation of Kronecker and the matrix multiplication.
The expectation is taken over
\begin{equation}
    \{\beta_{lm}, \phi_{\text{rx}}^{(lm)}, \theta_{\text{rx}}^{(lm)},\phi_{\text{tx}}^{(lm)}, \theta_{\text{tx}}^{(lm)}\}_{m = 1}^{20}.
\end{equation}
Thus, the CDL-A to C channel models are approximations of
\begin{equation}
    \bm{h} \sim \mathcal{N}_{\mathbb{C}}(\bm{0},\sum_{l=1}^L \bm{C}_l).
\end{equation}
The CDL-D and CDL-E additionally contain a \ac{LOS} path with uniformly distributed phase. Thus, we conclude that channel realizations from CDL models follow a near-Gaussian distribution. For models A to C, the approximation error arises from the central limit theorem, while for models D and E, there is an additional contribution to the error from the \ac{LOS} path.

\subsection{Other Link-Level Models}

Next to the 3GPP TDL and CDL models, there are several other link-level channel models. To our knowledge, they all exhibit similar characteristics and follow a (near-)Gaussian distribution. Examples are the (extended) pedestrian and vehicular link-level models \cite{3gppEPA}, as well as the TGn models A to F in the IEEE 802.11 standard for Wi-Fi channels \cite{TGn}. 

\subsection{Possible Gaussianity of Scenario-Level Channel Models}
It should be mentioned that scenario-level channel simulations can also lead to Gaussian channel characteristics. Particularly, the line of argumentation in Section \ref{sec:cdl} involves the central limit theorem for the inner summation in \eqref{eq:cdl_mimo}. However, if the number $L$ of paths in \eqref{eq:cdl_mimo} is large and all path characteristics are drawn \ac{i.i.d.} for all channel realizations, the channel itself is Gaussian, although considering varying path parameters. We observed, e.g., approximate Gaussian channel characteristics in pure \ac{NLOS} scenarios in QuaDRiGa. However, since this line of argumentation requires the same path characteristics for channel realizations at different locations in the scenario, it is a rather unrealistic exception.

\section{Experimental Section}

One of the key consequences of having Gaussian channels is that, for many physical-layer applications, the optimal method is provably linear and easily obtainable from the dataset that is assumed to be given when training \ac{ML} models. Thus, there is no need for \ac{ML} since we can simply compute the optimal method using classical schemes. We demonstrate this for the examples of \ac{CSI} compression, channel estimation, and \ac{ML}-aided channel modeling. Our code is publicly available.\footnote{\url{https://github.com/beneboeck/wireless-chan-mod4ml}}

\begin{figure}[t]
\NoHyper
    \centering
\includegraphics{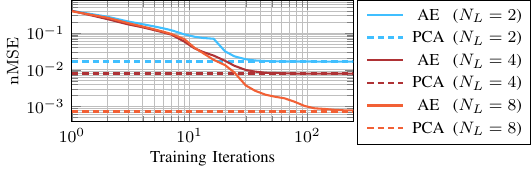}
\vspace{-0.6cm}
\caption{$\mathrm{nMSE}$ over training iterations on the TDL-D dataset.}
\label{fig:nmse_iterations_tdl}
	\vspace{-0.4cm}
\endNoHyper
\end{figure}

\begin{figure}[t]
\NoHyper
    \centering
\includegraphics{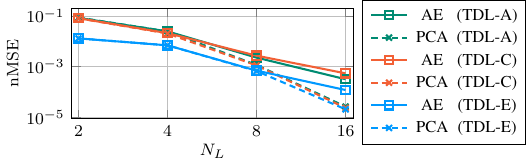}
   	\vspace{-0.6cm}
\caption{$\mathrm{nMSE}$ over $N_L$ for different link-level channel models.}
\label{fig:nmse_nl_compression}
	\vspace{-0.6cm}
\endNoHyper
\end{figure}

\subsection{CSI Compression}
\label{sec:csi_compression}
In \ac{FDD} systems, feeding back \ac{CSI} from the user to the \ac{BS} is essential for efficient communication \cite{love2008}. To do so without unnecessary overhead, one needs to compress \ac{CSI} at the user side via an encoder. The \ac{BS} then employs a decoder to reconstruct the fed-back signal. This compression can be done using an \ac{ML}-based \ac{AE} trained on a dataset of channel realizations $\mathcal{H} = \{\bm{h}_i\}_{i=1}^{N_t}$ \cite{wen2018}. The overall objective for training is
\begin{equation}
    \min_{\bm{\theta}, \bm{\phi}} \E_{p(\bm{h})}[\|\bm{h} - D_{\bm{\theta}}(E_{\bm{\phi}}(\bm{h}))\|_2^2]
\end{equation}
with $\bm{\theta}$-parameterized decoder $D_{\bm{\theta}}(\cdot)$ and $\bm{\phi}$-parameterized encoder $E_{\bm{\phi}}(\cdot)$ that maps $\bm{h}$ to a lower dimensional space $\mathbb{R}^{N_L}$ with predefined dimension $N_L$. Moreover, $\E_{p(\bm{h})}[\cdot]$ is approximated by means of $\mathcal{H}$. From standard literature about the \ac{PCA}, we know that when $\bm{h}$ is Gaussian, the optimal en- and decoder $E_{\bm{\phi}^*}(\cdot)$ and $D_{\bm{\theta}^*}(\cdot)$ are given by
\begin{equation}
    \label{eq:pca}
    \begin{aligned}
    E_{\bm{\phi}^*}(\bm{h}) = \bm{P}^{\operatorname{H}}\bm{h},\ \ D_{\bm{\theta}^*}(\bm{z}) = \bm{P}\bm{z}
    \end{aligned}
\end{equation}
with $\bm{P}$ containing the eigenvectors corresponding to the $N_L/2$ largest eigenvalues of the channel covariance matrix.\footnote{The number of eigenvectors is $N_L/2$ as we have a complex-valued compressed signal and, thus, its degree of freedom is twice its dimension.} Additionally, as we assume $\mathcal{H}$ to be given, the channel covariance matrix can be well estimated by the simple sample covariance 
\begin{equation}
\label{eq:scm}
    \hat{\bm{C}} = \frac{1}{N_t} \sum_{i=1}^{N_t} \bm{h}_i \bm{h}_i^{\operatorname{H}}.
\end{equation} 

\begin{table}[t]
\vspace{0.2cm}
    \centering
    \caption{\footnotesize $\mathrm{nMSE}$ for various link-level channel models ($N_L = 8$).}
    \label{tab:tdl}
\begin{tabular}{cccccc}
\toprule
\ & \footnotesize TDL-A & \footnotesize TDL-B & \footnotesize TDL-C & \footnotesize TDL-D & \footnotesize TDL-E 
\\
\midrule
\footnotesize AE & \footnotesize 0.00226 & \footnotesize 0.00145 & \footnotesize 0.00269 & \footnotesize 0.00075 & \footnotesize 0.00069 \\
\footnotesize PCA & \textbf{0.00115} & \textbf{0.00111} & \textbf{0.00111} & \textbf{0.00073} & \textbf{0.00067}\\
\bottomrule
\end{tabular}
\vspace{-0.5cm}
\end{table}

\begin{figure}[t]
\NoHyper
    \centering
\includegraphics{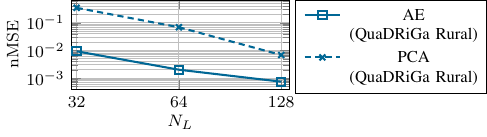}
   \vspace{-0.2cm}
\caption{$\mathrm{nMSE}$ over $N_L$ for the scenario-level QuaDRiGa rural dataset.}
\label{fig:nmse_quadriga_compression}
\endNoHyper
\end{figure}

Fig. \ref{fig:nmse_iterations_tdl} and \ref{fig:nmse_nl_compression} as well as Table \ref{tab:tdl} demonstrate the optimality of the \ac{PCA} for \ac{CSI} compression when using the TDL channel models. Specifically, we generate $60\,000$ \ac{OFDM} training channels with $60$kHz subcarrier spacing, $800$Hz maximal Doppler shift, $48$ subcarriers, $14$ time symbols, $0.25$ms overall duration, and $30$ ns delay spread with all TDL channel models, respectively. We normalize each dataset such that $\E[\|\bm{h}\|_2^2] = 48 \cdot 14$, where $\bm{h}$ is the vectorized \ac{OFDM} channel. The \ac{AE} structure is the same as that used in \cite{bock2025physicsinformed} (cf. \cite[Appendix E]{bock2025physicsinformed}). The performance metric is
\begin{equation}
\label{eq:nmse_def}
    \mathrm{nMSE} = \frac{1}{N_{v}N} \sum_{n=1}^{N_{v}} \|\bm{h}_n - \hat{\bm{h}}_n\|_2^2
\end{equation} 
with $\hat{\bm{h}}_n$ being the reconstructed channel of $\bm{h}_n$, test dataset size $N_v$  and channel dimension $N$. 
Fig. \ref{fig:nmse_iterations_tdl} shows the performance of the \ac{AE} during training on a validation dataset of $10\,000$ channels for different latent dimensions $N_L$. As a comparison, we also plot the performance of the \ac{PCA} according to \eqref{eq:pca}. While the \ac{AE}s improve over training, they all saturate at the performance of the linear \ac{PCA}. Fig. \ref{fig:nmse_nl_compression} illustrates the performance of the \ac{AE} against the \ac{PCA} for the channel models TDL-A, C, and E for different $N_L$. It can be seen that the \ac{AE} never outperforms the \ac{PCA}. Table \ref{tab:tdl} shows the same for all TDL models but with fixed $N_L = 8$. We conclude that when evaluating \ac{AE}s on TDL channel models, one essentially tests how well the \ac{AE} can approximate the linear mapping in \eqref{eq:pca}, which can also be directly obtained from the training dataset $\mathcal{H}$.
Fig. \ref{fig:nmse_quadriga_compression} shows the performance on a scenario-level QuaDRiGa \ac{LOS} rural scene, where we used the same configurations as for the TDL models.\footnote{For more information about the dataset, we refer to \cite[Appendix D]{bock2025physicsinformed}.} We see that the \ac{AE} outperforms the \ac{PCA} resulting from the non-Gaussianity of the scenario-level QuaDRiGa rural channel model.

\begin{figure}[t]
\NoHyper
\vspace{-0.2cm}
    \centering
\includegraphics{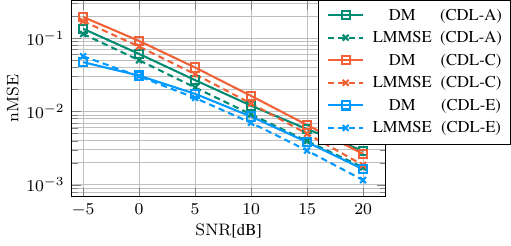}
   \vspace{-0.2cm}
\caption{$\mathrm{nMSE}$ over $\mathrm{SNR}$ in dB for different link-level channel models.}
\label{fig:nmse_snr_estimation_linklevel}
	\vspace{-0.6cm}
\endNoHyper
\end{figure}

\subsection{Channel Estimation}
\label{sec:channel_estimation}
Accurate knowledge of \ac{CSI} and, thus, channel estimation is of key importance in \ac{MIMO} systems \cite{massivemimobook}.  In general, the goal of channel estimation is to minimize the \ac{MSE} between the actual channel $\bm{h}$ and an estimate $\hat{\bm{h}}(\bm{y})$ given a potentially compressed and noisy observation $\bm{y} = \bm{A}\bm{h} + \bm{n}$ with \ac{AWGN} $\bm{n} \sim \mathcal{N}_{\mathbb{C}}(\bm{0}, \sigma_n^2 \operatorname{\mathbf{I}})$ and measurement matrix $\bm{A}$. Typically, $\bm{A}$ and $\sigma_n^2$ are assumed to be known. The \ac{MSE}-optimal estimator is the \ac{CME}, i.e., $\hat{\bm{h}}(\bm{y}) = \E[\bm{h}|\bm{y}]$, which is, for a non-Gaussian $\bm{h}$, typically non-linear. However, for a Gaussian $\bm{h}$, the \ac{CME} reduces to the \ac{LMMSE} estimator
\begin{equation}
    \hat{\bm{h}}(\bm{y}) = \bm{\mu}_{\bm{h}} + \bm{C}_{\bm{h}}\bm{A}^{\operatorname{H}}(\bm{A}\bm{C}_{\bm{h}}\bm{A}^{\operatorname{H}} + \sigma_n^2 \operatorname{\mathbf{I}})^{-1}(\bm{y} - \bm{\mu}_{\bm{h}})
\end{equation}
with $\bm{h} \sim \mathcal{N}_{\mathbb{C}}(\bm{\mu}_{\bm{h}}, \bm{C}_{\bm{h}})$. When training a \ac{ML}-based channel estimator, we either assume a large training dataset $\mathcal{D} = \{\bm{y}_i, \bm{h}_i\}_{i=1}^{N_t}$ or $\mathcal{H} = \{\bm{h}_i\}_{i=1}^{N_t}$ to be given. This data can be used to estimate the channel mean and covariance as in \eqref{eq:scm}.  

Fig. \ref{fig:nmse_snr_estimation_linklevel} and Table \ref{tab:estimation} demonstrate the (approximate) optimality of the \ac{LMMSE} for link-level channel models. In particular, we generate $60\,000$ \ac{MIMO} training channels with $16$ transmit and $8$ receive antennas with all CDL channel models, respectively. On both sides, we use a \ac{ULA} with $\lambda/2$ antenna spacing. The center frequency is set to $3.5$GHz. We normalize the whole dataset such that $\E[\|\bm{h}\|_2^2] = 16 \cdot 8$, where $\bm{h}$ is the vectorized \ac{MIMO} channel. As a \ac{ML}-based estimator, we utilize the \ac{DM}-based estimator from \cite{feslwcl}, which is known to be asymptotically optimal in case $\bm{A} = \operatorname{\mathbf{I}}$ \cite{fesl2025on}. We assume $\bm{A} = \operatorname{\mathbf{I}}$ throughout all simulations. Fig. \ref{fig:nmse_snr_estimation_linklevel} shows the estimation performance in $\mathrm{nMSE}$ (cf. \eqref{eq:nmse_def}) over the \ac{SNR} defined as $\mathrm{SNR} = \E[\|\bm{h}\|_2^2]/(16 \cdot 8 \cdot \sigma_n^2) = 1/\sigma_n^2$ for the CDL-A, C and E model. For the \ac{NLOS} models A and C, we see that the \ac{LMMSE} outperforms the \ac{DM} over the whole \ac{SNR} range. This validates the approximation of the central limit theorem in \eqref{eq:central_limit}. As CDL-E contains a further \ac{LOS} path, it is not perfectly Gaussian, but near-Gaussian. Consequently, we can see that the \ac{DM} slightly outperforms the \ac{LMMSE} in the low SNR regime. However, the \ac{LMMSE} performs better for all other \ac{SNR} values, rendering the Gaussian approximation to be highly accurate. Table \ref{tab:estimation} confirms these insights by presenting the $\mathrm{nMSE}$ for all CDL channel models at $10$dB \ac{SNR}. 

\begin{figure}[t]
\NoHyper
    \centering
\includegraphics{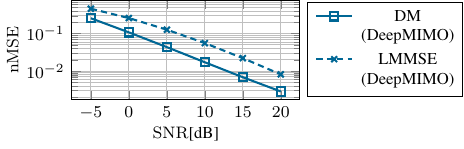}
   \vspace{-0.2cm}
\caption{$\mathrm{nMSE}$ over $\mathrm{SNR}$ in dB for the DeepMIMO Boston scenario.}
\label{fig:nmse_snr_deepmimo}
\endNoHyper
\end{figure}

Fig. \ref{fig:nmse_snr_deepmimo} shows the comparison between the \ac{DM} and the \ac{LMMSE} for the scenario-level DeepMIMO channel simulator based on the Boston scenario \cite{Alkhateeb2019}.\footnote{For more information about the dataset, we refer to \cite[Appendix D]{bock2025physicsinformed}.} We see that the \ac{ML}-based \ac{DM} estimator easily outperforms the \ac{LMMSE} resulting from the non-Gaussianity of the DeepMIMO Boston channel data.

\begin{table}[t]
\vspace{-0.2cm}
    \centering
    \caption{\footnotesize $\mathrm{nMSE}$ for various link-level channel models ($\mathrm{SNR} = 10$dB).}
    \label{tab:estimation}
\begin{tabular}{cccccc}
\toprule
\ & \footnotesize CDL-A & \footnotesize CDL-B & \footnotesize CDL-C & \footnotesize CDL-D & \footnotesize CDL-E 
\\
\midrule
\footnotesize DM & \footnotesize 0.01204 & \footnotesize 0.01928 & \footnotesize 0.01637 & \footnotesize 0.00657 & \footnotesize 0.00856 \\
\footnotesize LMMSE & \textbf{0.00915} & \textbf{0.01555} & \textbf{0.01288} & \textbf{0.00525} & \textbf{0.00703}\\
\bottomrule
\end{tabular}
\vspace{-0.5cm}
\end{table}

\begin{figure*}[t]
\NoHyper
    \centering
    \includegraphics{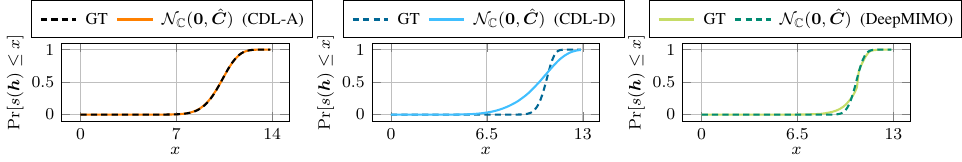}
   \vspace{-0.3cm}
\caption{CDFs of the spectral efficiency for different channel models ($\sigma_n^2$ is chosen such that $\mathrm{SNR} = 20$dB).}
\label{fig:cdf}
	\vspace{-0.6cm}
\endNoHyper
\end{figure*}

\begin{figure}[t]
\vspace{-0.35cm}
    \centering
    \includegraphics{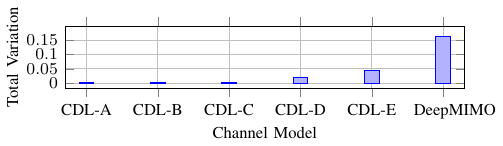}
   \vspace{-0.2cm}
\caption{Total variation for different channel models.}
\label{fig:tv}
	\vspace{-0.65cm}
\end{figure}

\subsection{\ac{ML}-aided Channel Modeling}

While channel models provide training data for \ac{ML}-based methods, there is also ongoing research on directly using \ac{ML} to learn a proper site-specific channel model \cite{Yang2019, bock2025physicsinformed}. Given a dataset of channel realizations $\mathcal{H} = \{\bm{h}_i\}_{i=1}^{N_t}$, all \ac{ML}-based channel models have in common that they aim to learn a generator $\mathcal{G}_{\bm{\theta}}(\cdot)$ that maps realizations $\bm{z}$ from a simplistic distribution (e.g., $\mathcal{N}_{\mathbb{C}}(\bm{0}, \operatorname{\mathbf{I}})$) to a channel realization. This mapping can either be fully deterministic \cite{Yang2019} or stochastic \cite{bock2025physicsinformed}. For $\bm{h}$ being Gaussian, the optimal generator $\mathcal{G}_{\bm{\theta}^*}(\cdot)$ is
\begin{equation}
    \label{eq:opt_generator}
    \mathcal{G}_{\bm{\theta}^*}(\bm{z}) = \bm{U}\sqrt{\bm{\Lambda}}\bm{z}
\end{equation}
with $\bm{z} \sim \mathcal{N}_{\mathbb{C}}(\bm{0}, \operatorname{\mathbf{I}})$ and $\bm{U}\bm{\Lambda}\bm{U}^{\operatorname{H}}$ being the eigenvalue decomposition of the channel covariance $\bm{C}_{\bm{h}}$.\footnote{Without loss of generality, we assume the channel to have a zero mean.} Equivalent to \ac{CSI} compression as well as channel estimation (cf. Section \ref{sec:csi_compression} and \ref{sec:channel_estimation}), we estimate the channel covariance using the sample covariance in \eqref{eq:scm} over $\mathcal{H}$. For validating the optimality of \eqref{eq:opt_generator}, we use the spectral efficiency analysis and codebook fingerprinting technique from \cite{baur2025}. In particular, we compare the \acp{CDF} of the spectral efficiency $s(\bm{h})$ when transmitting data via a noisy spatial system model $\bm{h} + \bm{n}$ ($\bm{n} \sim \mathcal{N}_{\mathbb{C}}(\bm{0}, \sigma_n^2 \operatorname{\mathbf{I}})$), i.e., $s(\bm{h}) = \log_2(1 + \|\bm{h}\|_2^2/\sigma_n^2)$.
We input channel realizations from the channel model at hand (GT) as well as channel realizations from \eqref{eq:opt_generator} in $s(\bm{h})$, respectively. We use a $16$-antenna \ac{ULA} with $\lambda/2$ spacing at the receiver and $1$ antenna at the transmitter. We normalize the whole dataset such that $\E[\|\bm{h}\|_2^2] = 16$. Fig. \ref{fig:cdf} shows the \ac{CDF} comparison for the three different channel models CDL-A, CDL-D, and the Boston DeepMIMO scenario. As the CDL-A model is an \ac{NLOS} link-level channel model, both \acp{CDF} perfectly coincide. The CDL-D model contains a \ac{LOS} path, which is reflected by the \acp{CDF} not perfectly overlapping. Interestingly, although the DeepMIMO Boston scenario is a scenario-level channel model with non-Gaussian characteristics, the spectral efficiency exhibits similar characteristics to those produced by \eqref{eq:opt_generator}. Since $s(\bm{h})$ only depends on the channel norm, this measure is not sufficient for validating the realism of generated samples. Therefore, we also evaluate the generation performance using the codebook fingerprinting method from \cite{baur2025}. We use the \ac{DFT} matrix as codebook and compute the total variation between the \acp{PMF} of the resulting indices. This is done by projecting channel realizations on the codebook entries and extracting the best-fitting index in absolute value. The resulting histogram is normalized and interpreted as \ac{PMF}, allowing a comparison using the total variation \cite{baur2025}. The results can be seen in Fig. \ref{fig:tv}. CDL-A to C models exhibit perfect Gaussianity. The CDL-D and CDL-E show a slightly larger total variation, while DeepMIMO has the largest one, validating its non-Gaussianity.
\section{Conclusion}

We critically reviewed standardized link-level channel models, a frequently used class of channel models in \ac{ML}-aided wireless communications. We discussed the drawbacks of the implicit assumptions when evaluating \ac{ML}-based methods using these models and demonstrated that for many physical-layer applications, classical signal processing outperforms \ac{ML} in the link-level perspective. We also discussed that the scenario-level perspective offers a solution to all the drawbacks of link-level simulations and renders \ac{ML} worthwhile. 

\section*{Acknowledgements}
The authors thank Moritz Hocher (\url{https://www.moritz-hocher.com/}) for creating the 3D scene in Fig.~\ref{fig:linklevelvescenariolevel}.
\bibliographystyle{IEEEtran}
\bibliography{references}

\end{document}